\documentclass[aps,preprint,nofootinbib]{revtex4}
\usepackage{graphicx}
\usepackage{amsmath,amssymb}
\usepackage{url}
\usepackage{color}
\usepackage{epstopdf}
\newcommand{\lsim}{\mathrel{\mathop{\kern 0pt \rlap
  {\raise.2ex\hbox{$<$}}}
  \lower.9ex\hbox{\kern-.190em $\sim$}}}
\newcommand{\gsim}{\mathrel{\mathop{\kern 0pt \rlap
  {\raise.2ex\hbox{$>$}}}
  \lower.9ex\hbox{\kern-.190em $\sim$}}}

\begin{document}
\title{Implication on Higgs invisible width in light of the new CDMS result}
\author{Kingman Cheung$^{1,2,3}$ and Tzu-Chiang Yuan$^4$}

\affiliation{
$^1$Division of Quantum Phases \& Devices, School of Physics, 
Konkuk university, Seoul 143-701, Korea \\
$^2$Department of Physics, National Tsing Hua University, 
Hsinchu 300
\\
$^3$Physics Division, National Center for Theoretical Sciences,
Hsinchu 300
\\
$^4$Institute of Physics, Academia Sinica, Nankang, Taipei 11529, Taiwan
}

\date{\today}

\begin{abstract}
With the assumption that the dominant diagram in supersymmetry 
for the spin-independent cross section of the dark matter particle 
is due to Higgs boson exchange, we obtain an upper limit on the
Higgs-dark-matter coupling based on the new result of the CDMSII 
Collaborations.  We then obtain an upper limit on the invisible 
width of the Higgs boson, numerically it is less than $20-120$ MeV
for $m_h \simeq 120 - 180$ GeV.
Implications for Higgs boson search are also discussed.
\end{abstract}
\maketitle

{\it Introduction.}
The presence of cold dark matter (CDM) in our Universe is now well established
by the very precise measurement of the cosmic microwave background radiation
in the Wilkinson Microwave Anisotropy Probe (WMAP) experiment \cite{wmap}.
A nominal $3\sigma$ range of the CDM relic density is
\[
 \Omega_{\rm CDM}\, h^2 = 0.105 \;^{+0.021}_{-0.030} \;,
\]
where $h$ is the Hubble constant in units of $100$ km/Mpc/s.

One of the most appealing and natural CDM particle candidates is provided by
supersymmetric models with $R$-parity conservation \cite{hooper}.
This $R$-parity  conservation ensures the stability of the lightest
supersymmetric particle (LSP) so that the LSP can be CDM.
The LSP is in general the lightest neutralino, a linear combination of
neutral electroweak (EW) gauginos and Higgsinos.
Since the LSP nature depends on its compositions,
its detection can vary a lot.

One of the most direct detection methods of the dark matter is through a set of
direct search experiments.  The CDMS is one of these.  
The dark matter particles move at a velocity relative to the detecting
materials.  It will recoil against the nucleons, and create a phonon-type
signal, which can be amplified by electronics.  
Just very recently the CDMSII finalized their search in Ref. \cite{cdms}.
When they opened the black box in their blind analysis,
they found two candidate events, which are
consistent with background fluctuation at a probability level of
about 23\%.  This exciting news already stimulates a large number of
activities \cite{recent} in a very short period of time.
Nevertheless, the signal is not conclusive.  The CDMS then improves 
upon the upper limit on the 
spin-independent cross section $\sigma^{\rm SI}_{\chi N}$ to
$3.8 \times 10^{-44} \; {\rm cm}^2$ for $m_\chi \approx 70 $ GeV.  
In the following, we use this new limit to put a new bound on the
Higgs-dark-matter coupling, which is then implied to an upper limit on
the Higgs boson invisible width.  This is the main result of this work.
Further implications for the Higgs boson search at colliders 
are also discussed.

{\it Direct Detection.}
The spin-independent cross section between the dark matter particle
(denoted by $\chi$ in the following) and the nucleon is given by
\begin{equation}
\sigma^{\rm SI}_{\chi N} = \frac{\mu_{\chi N}^2}{\pi}\; \left|
 G^N_s \right |^2 \;,
\end{equation}
where $\mu_{\chi N} = m_\chi m_N / ( m_\chi + m_N)$ is the reduced mass
between the dark matter particle and the nucleon, and 
\begin{equation}
G^N_s = \sum_{qq=u,d,s,c,b,t} \langle N | \bar q q | N \rangle \;
\left ( \frac{1}{2} \sum_{q}
  \frac{  g_{L \tilde{q} q} g_{R \tilde{q} q} }{ m^2_{\tilde{q} } }
- \frac{  g_{h\chi \chi} g_{h q q } }{ m^2_h } \right  )
\end{equation}
Suppose the squarks are heavy, like that in split SUSY, the dominant
diagram is the Higgs boson exchange diagram.  The upper limit
on the spin-independent cross section can imply an upper limit
for the dark matter-Higgs coupling, which is more or less model
independent.  We shall ignore the squark exchange in the following.  

\begin{figure}[th!]
\centering
\includegraphics[width=3in]{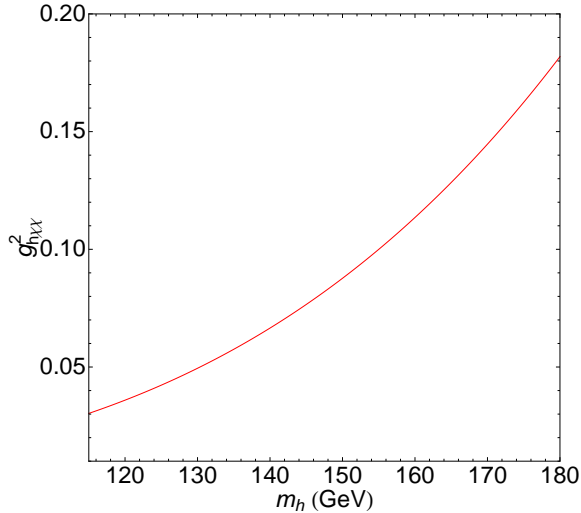}
\caption{Contour of $\sigma^{\rm SI}_{\chi N} = 3.8 \times 10^{-44} \; {\rm cm}^2$ 
as function of $m_h^2$ and $g^2_{h\chi\chi}$.  The curve is also 
the upper limit on the coupling $g_{h\chi\chi}$.
\label{fig}
}
\end{figure}

Default values of the parameters used, e.g. in DarkSUSY \cite{darksusy}
are (with $\langle N | \bar q q | N \rangle = f^T_{Tq} m_N/m_q$)
\begin{eqnarray}
&&f^p_{Tu} = 0.023,\;\; f^p_{Td} = 0.034,\;\; f^p_{Ts} = 0.14,\;\;
f^p_{Tc} = f^p_{Tb} = f^p_{Tt} =  0.0595 \;, \nonumber \\
&&f^n_{Tu} = 0.019,\;\; f^n_{Td} = 0.041,\;\; f^n_{Ts} = 0.14,\;\;
f^n_{Tc} = f^n_{Tb} = f^n_{Tt} =  0.0592 \;, 
\end{eqnarray}
We take the average between proton and neutron for $\sigma^{\rm SI}_{\chi N}$.
Note that the $m_q$ dependence in the Yukawa coupling $g_{h qq}$ 
will be cancelled by the $m_q$ dependence in 
$\langle N | \bar q q | N \rangle $.
Taking the average between proton and neutron the value of $G^N_s$ is
\begin{equation}
- G^N_s \simeq g_{h\chi\chi} \frac{g m_p}{2 m_W} \frac{1}{m^2_h} (0.3766) \;.
\end{equation}
For $m_\chi \sim O(100)$ GeV $\mu_{\chi p} \approx m_p$.  The 
spin-independent cross section is
\begin{equation}
\sigma^{\rm SI}_{\chi N} \approx \frac{ g^2 m_p^4}{4 \pi m_W^2}\, \frac{1}{m_h^4}\,
 g^2_{h\chi \chi} (0.3766)^2 \;.
\end{equation}
We can take $m_h = 115 - 180$ GeV for most general SUSY models and
use the new CDMSII limit $\sigma^{\rm SI}_{\chi N} < 3.8 \times 10^{-44} \;
{\rm cm}^2$.  We can obtain an upper limit on the Higgs-dark-matter coupling
\begin{equation}
\label{limit}
g^2_{h\chi \chi} \alt 0.03 -  0.18
\end{equation}
for $m_h = 115 - 180 $ GeV, which takes into account the current Higgs mass
lower limit and the predictions on the upper limit of the lightest Higgs boson
in most SUSY models.
We show in Fig.~\ref{fig} the upper limit on $g_{h\chi\chi}^2$ versus $m_h$
from the result $\sigma^{\rm SI}_{\chi N} < 3.8 \times 10^{-44}\;{\rm cm}^2$.

Since the recoil energies of the two signal events recorded by the CDMSII 
were $10-15$ keV, the mass of the dark matter particle is at most around 
100 GeV.  The lower limit on the the $\sigma^{\rm SI}_{\chi N}$ around 100 GeV
is all about $3.8 - 5 \times 10^{-44}\; {\rm cm}^2$, thus the lower limits
on $g_{h\chi\chi}$ for $m_\chi$ around 100 GeV are about the same.  

If the Higgs boson is heavy enough, say $150 - 180$ GeV and the dark matter
particle is less than about 75 GeV, the Higgs boson can decay into 
a pair of dark matter particles.  It will give rise to invisible width
of the Higgs boson in collider experiments.  Based on the upper limit 
that we obtain in Eq. (\ref{limit}), we can derive an upper limit on
the Higgs invisible width
\begin{equation}
\Gamma_{\rm inv} = \frac{g_{h \chi\chi}^2 m_h}{16 \pi} \left( 1 - 
\frac{4 m_\chi^2}{m_h^2} \right )^{3/2} \;,
\end{equation}
where we assume dark matter particle is Majorana. Numerically, using the
limit in Eq. (\ref{limit}) we obtain the upper limit on the invisible width
of the Higgs boson for $m_h \approx 180$ GeV
\begin{equation}
\label{wid}
\Gamma_{\rm inv} < 20 - 120 \; {\rm MeV} \;.
\end{equation}
Note that the signal events, if interpreted as signals, indicate a cross
section of order $O(10^{-44})\; {\rm cm}^2$, which in turn gives rise to
an invisible width of order $10 - 50$ MeV.

{\it Implications for Higgs search.}
We assume the other SUSY particles are heavy such that the only
SUSY particle that the lightest Higgs boson can decay into is the
lightest neutralino -- the dark matter particle that we denote by
$\chi$.  The mass range of the Higgs boson that we consider in this work
is $120 - 180 $ GeV, while the dark matter particle is less than $m_h/2$.
The recoil energies of the signal events recorded by the CDMSII
tell us that the mass of the dark matter particle is around 100 GeV.
The dominant decay mode of the SM-like Higgs boson is either $b\bar b$ or
the $WW^*$ in the range considered.  
The total decay width of the SM Higgs boson is about 
a few MeV at $m_h = 120$ GeV and sharply rises to 1 GeV at $m_h = 180$ GeV,
because of the opening of the $WW^*$ model. The SM-like
lightest CP-even Higgs boson in SUSY models has a similar decay width.
Therefore, the limit obtained in Eq. (\ref{wid}) is most relevant for 
the lower mass range, where the $WW$ or $WW^*$ mode is not yet open.
If we take the invisible width to be of order of $O(10)$ MeV at $m_h = 120-140$ 
GeV, then the invisible decay of the Higgs boson can be dominant. This will 
have nontrivial effects on the search for the intermediate Higgs boson
at colliders.

The search at the Tevatron is mostly based on associated $Wh, Zh$ production,
followed by leptonic decays of the $W$ or $Z$ and the $b\bar b$ or $WW^*$ 
mode of the Higgs boson. If the invisible decay becomes significant, then
the fraction into $b\bar b$ would be less, such that the search would be
more difficult.  Nevertheless, for $m_h > 140$ GeV the $WW^*$ mode opens
and the invisible mode is less significant. 
The Higgs boson search at the LHC for $m_h = 120-140$ GeV 
is based on $gg\to h \to \gamma\gamma$.  It may be worsened if the
invisible mode becomes significant.
One can search for invisible Higgs boson quite efficiently at LEPII,
because the Higgs boson mass can be reconstructed using the recoil mass
of the visible decay products of the $Z$ boson.
The search for invisible mode at LEPII has been performed. The lower
limit on the Higgs boson mass is almost the same as the SM lower limit,
provided that the Higgs boson decays 100\% invisibly.  

In summary, we have used the new CDMSII result on the spin-independent cross
section to obtain an upper limit on the Higgs-dark-matter coupling,
assuming the Higgs boson exchange is the dominant mechanism. We then 
derive an upper limit on the invisible width of the Higgs boson.
If we take the two signal events seriously, the cross section of the order 
of $10^{-44}\; {\rm cm}^2$ implies an invisible width of order 10 MeV
for the Higgs boson.  Such an invisible width would have significant
effects on search for intermediate Higgs boson of mass $120-140$ GeV.

\section*{Acknowledgments}
The work was supported by the NSC of Taiwan, 
the Boost Project of NTHU, and the WCU program through the KOSEF funded
by the MEST (R31-2008-000-10057-0). 



\begin{thebibliography}{99}
\bibitem{wmap}
D.~N.~Spergel {\it et al.}  [WMAP Collaboration],
  Astrophys.\ J.\ Suppl.\  {\bf 170}, 377 (2007)
  [arXiv:astro-ph/0603449].

\bibitem{hooper}
G.~Bertone, D.~Hooper, and J.~Silk,
  Phys.\ Rept.\  {\bf 405}, 279 (2005)
  [arXiv:hep-ph/0404175].

\bibitem{cdms}
Z.~Ahmed {\it et al.}  [The CDMS-II Collaboration and CDMS-II
                  Collaboration],
  arXiv:0912.3592 [astro-ph.CO].

\bibitem{recent}
 M.~Kadastik, K.~Kannike, A.~Racioppi and M.~Raidal,
  arXiv:0912.3797 [hep-ph].

A.~Bottino, F.~Donato, N.~Fornengo and S.~Scopel,
  arXiv:0912.4025 [hep-ph].

D.~Feldman, Z.~Liu and P.~Nath,
  arXiv:0912.4217 [hep-ph].

M.~Ibe and T.~T.~Yanagida,
  arXiv:0912.4221 [hep-ph].

J.~Kopp, T.~Schwetz and J.~Zupan,
  arXiv:0912.4264 [hep-ph].

R.~Allahverdi, B.~Dutta and Y.~Santoso,
  arXiv:0912.4329 [hep-ph].

M.~Endo, S.~Shirai and K.~Yonekura,
  arXiv:0912.4484 [hep-ph].

 M.~Holmes and B.~D.~Nelson,
  arXiv:0912.4507 [hep-ph].

Q.~H.~Cao, I.~Low and G.~Shaughnessy,
  arXiv:0912.4510 [hep-ph].

  Q.~H.~Cao, C.~R.~Chen, C.~S.~Li and H.~Zhang,
  arXiv:0912.4511 [hep-ph].

\bibitem{darksusy}
P.~Gondolo, J.~Edsjo, P.~Ullio, L.~Bergstrom, M.~Schelke and E.~A.~Baltz,
  JCAP {\bf 0407}, 008 (2004)
  [arXiv:astro-ph/0406204].


\end{thebibliography}
\end{document}